\documentclass[sigplan,10pt]{acmart}

\renewcommand\footnotetextcopyrightpermission[1]{}

\usepackage[utf8]{inputenc}
\usepackage{url}
\usepackage{xurl}
\usepackage{todonotes}

\usepackage{flushend}

\begin{document}
\title{Architecting a reliable quantum operating system: microkernel, message passing and supercomputing}

\author{Alexandru Paler}
\affiliation{%
  \institution{Aalto University, Finland}
  }
\email{alexandru.paler@aalto.fi}

\begin{abstract}
A quantum operating system (QCOS) is a classic software running on classic hardware. The QCOS is preparing, starting, controlling and managing quantum computations. The reliable execution of fault-tolerant quantum computations will require the QCOS to be as reliable and fault-tolerant as the computation itself. In the following, we discuss why a QCOS should be architected according to the following principles: 1) using a microkernel; 2) the components are working in an aggregated, non-stacked manner and communicate by message passing; 3) the components are executed by default on supercomputers, unless there are very good reasons not to. These principles can guarantee that the execution of error-corrected, fault-tolerant quantum computation is not vulnerable to the failures of the QCOS.
\end{abstract}

\maketitle
\pagestyle{empty}

\section{Introduction}

Technology roadmaps (e.g.~\cite{quantinuumQuantinuumAccelerates, riverlaneIntroducingRiverlanes}), envision the early 2030s as the moment when large scale quantum computers might be operating. To have a QCOS ready by that deadline, its components have to be as small as possible, well defined, working and integrated\footnote{This is a shorter and updated version of the manuscript published in~\cite{paler2020aggregated}}. Therefore, this is not to open a Tanenbaum -- Torvalds-like debate, but to use parts of the experience of NASA's Apollo programme when building the first QCOS. In a nutshell, the Apollo approach~\cite{haigh2018hey} is: 1) meeting a fixed deadline means to work backward to identify the points by which sub-systems have to be ready and integrated; 2) given a choice between two technologically workable ways to do something, take the better, proven and more expensive way. We argue that the better, proven and more expensive way are supercomputers running a QCOS built on top of a microkernel. 

Practical quantum computations will require millions of qubits (e.g.~\cite{campbell2017roads, gidney2021factor}), such that distributed quantum computers have been proposed~\cite{van2016path}. At the same time, it has been shown that a QCOS could use Grover's algorithm to speed up classical OS functions like  scheduling~\cite{britt2017quantum}. Although any OS could benefit from quadratic speedups, qubits are such a scarce resource that it would be better to use them for quantum chemistry, for example. 

It is expected that a practical, quantum error-corrected computation will run between a few hours~\cite{babbush2018encoding} and a few months. Achieving extremely close to 100\% availability for a very large quantum computation requires the QCOS to be also 100\% available, and this is a significant challenge due to failures affecting some potentially small portions of the classical hardware or QCOS\cite{rojas2019analyzing}.

The fault-tolerance of the QCOS plays a significant role in evaluating the total reliability of an arbitrary quantum computation. In general, it is common for systems to fail even when every component is correct and seems secure~\cite{bellovin2018big}. It would be ironic for the quantum software to be the weakest point in the fault-tolerance of a error-corrected quantum computation.

\begin{figure}[t!]
    \centering
    \includegraphics[width=0.8\columnwidth]{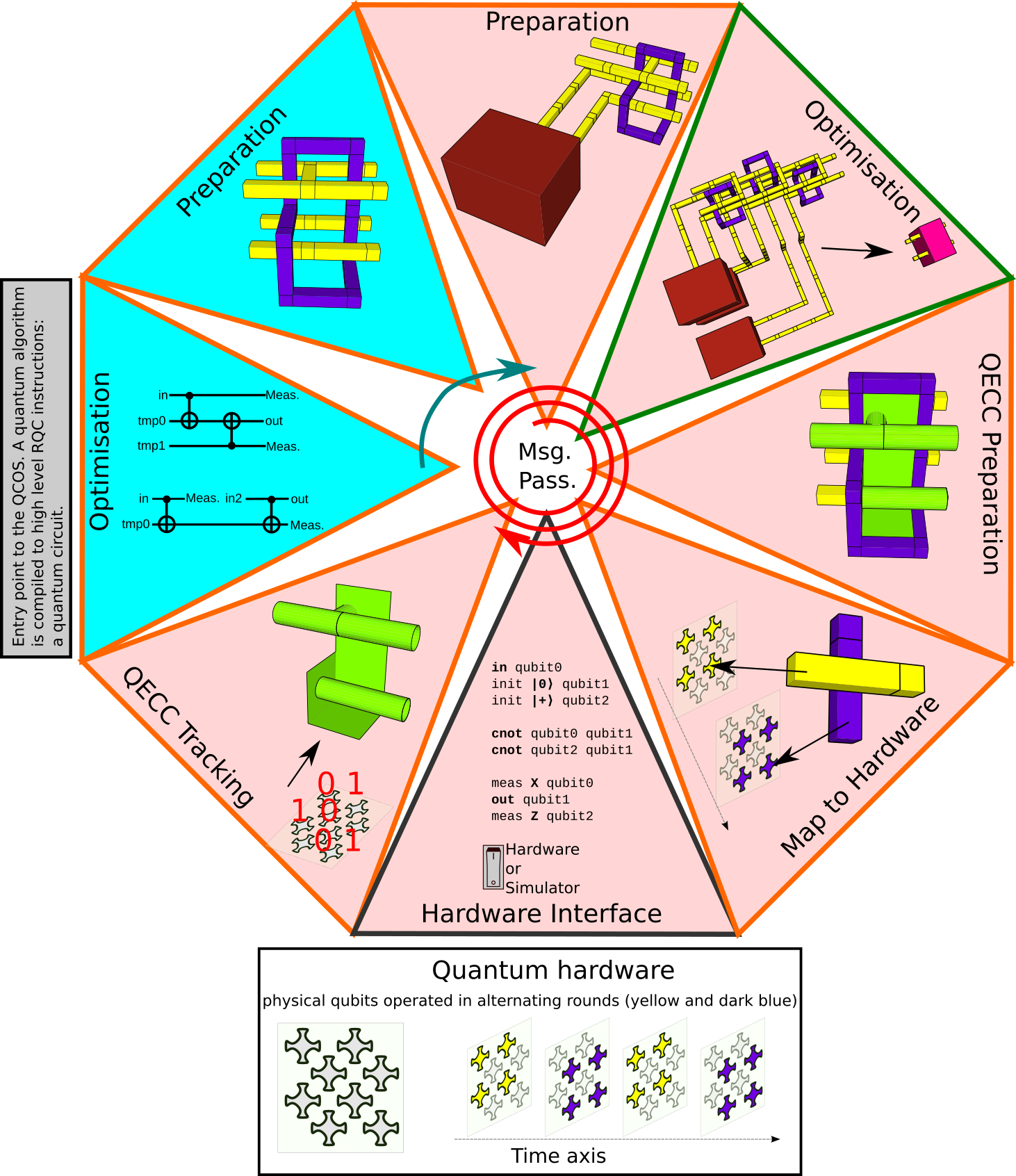}
    \caption{Architecture and interaction diagram of the QCOS. Each triangle is a QCOS component. The arrows indicate a non-strict execution order of the components. To execute a quantum algorithm, the first part of the QCOS is offline (blue arrow and triangles), after which online preparations, optimisations etc. are performed in a loop (red spiral and triangles). The component pictograms represent elements of quantum circuits protected by the braided surface quantum error-correcting code~\cite{van2013blueprint}.}
    \label{fig:arch}
\end{figure}

We propose a QCOS architecture by building on the principles of fault-tolerant classical operating systems~\cite{denning1976fault}(process isolation, resource control, decision verification and error recovery). The goal of those principles is \emph{error confinement} -- limit the risk that errors do much damage before being detected. To this end, herein we argue that the QCOS should be built: 1) using a microkernel; 2) from aggregated/distributed message-passing components; 3) such that components are executed on a supercomputer by default. In the following sections we motivate our architecture.

\section{Building on top of a microkernel}

Quantum computer control software has been proposed having layered, stacked and densely interacting software components (e.g.~\cite{fu2018microarchitecture}). This is a trait of \emph{almost-monolithic architectures}. Monolithic kernels do not exhibit high levels of fault-tolerance (e.g. a driver crash can panic the kernel and stop the entire system).

In order to guarantee system reliability~\cite{avizienis2004basic}, such as high availability and high fault tolerance, our proposal is to depart from monolithic architectures and to use distributed microkernels (so-called splitkernels~\cite{shan2018legoos}) in order to increase the QCOS fault-tolerance.

Microkernels are motivated by the desire to minimise the exposure to bugs~\cite{klein2014comprehensive}, and thus to system failures. Microkernels are slow, difficult to implement and debug, and because of these reasons, for example, Linux uses a monolithic kernel. Nevertheless, microkernels can be formally verified -- showing that the kernel implementation is consistent with its specification and does not include bugs~\cite{klein2014comprehensive}. Formal verification is great advantage, and would place a QCOS on par with the importance of OS used in aerospace and defense.

A microkernel assumes that the QCOS is light and the components are applications, whereas in a splitkernel the components are parts of the kernel, and have high execution priority. A splitkernel for classical OS has been proposed by~\cite{shan2018legoos}, and  the overall system fault-tolerance has increased, while the performance did not drop significantly -- we increase the fault-tolerance by using micro- and split-kernels at the same time.

\section{Message-passing components}

The QCOS will be started each time a quantum algorithm is executed. The simplest QCOS architecture will have a loosely coupled and distributed architecture, which will include components for circuit compilation, optimisation, error-correction, decoding etc. These components are running on top of the microkernel.

Our QCOS has an aggregated architecture, meaning that there is no top-down work-flow, and all components operate at the same level (Figure~\ref{fig:arch}). We use message passing, because it has been successfully used to engineer very reliable distributed systems (e.g. Erlang), it is compatible with the splitkernel and the supercomputing hardware discussed in the next section. Therefore, the QCOS components and the splitkernel should use message-passing (e.g. MPI), which is a natural communication strategy in supercomputing.

The unique character of the QCOS is given by its online (red) components, which are executed  to take corrective measures in a real-time, low-latency and resilient manner in order to maintain computational fault-tolerance. The advance being proposed here is the scalable and resilient feedback-loop between the high level quantum algorithm and the quantum computer. The microkernel includes a scheduler which is responsible for the just in time execution scheduling of the components, and for dynamically updating the execution priorities of the components.

In Figure~\ref{fig:arch} there is a QCOS boot phase (offline, blue triangles), and a QCOS execution phase (online, red triangles). During boot, the quantum algorithm is compiled and optimised in the form of a quantum circuit. The next step is to prepare the error-corrected quantum circuit offline. Offline compilation, optimisation and error-correction are very complex and may take a lot of time. The booting phase is an abstraction of the entire procedure until the actual execution of the quantum algorithm is started.

During the QCOS’ execution, the circuits are prepared (compiled) and optimised just in time. The error-corrected representation of the computations is generated by the QECC preparation component. The error-corrected form is then translated to quantum hardware instructions by the Hardware Mapper. This component is sending instructions to the quantum computer through the hardware interface. The QCOS schedules instructions into discrete rounds. For example, in Figure~\ref{fig:arch}, in the quantum hardware box, there are yellow and magenta rounds, and green crosses represent the hardware qubits. Finally, the QECC Tracking component is responsible for decoding and Pauli frame tracking~\cite{paler2014software}.

Real-time error decoding is a topic of intense research, because the QCOS has to face immense data rates generated by the millions of qubits~\cite{battistel2023real}. Quantum measurements are probabilistic, and instruction execution generates a probabilistic binary measurement result. The feedback for correcting the computations is formed by the measurement results collected through the hardware interface. Error-correction tracking uses the feedback stream: error syndromes are computed, and the execution of the quantum computation is dynamically adapted in case computational corrections are needed.

The online (red) components are not executed in a strict order (e.g. top-down like in standard software stack), but just in time, depending on the needs of the computation. For example, the hardware mapping component might raise an interrupt that the optimized circuit does not fit in the available hardware, or the QECC tracking component raises an interrupt that the circuit has to be re-compiled because of some necessary corrective gates. This is similar to a classical OS, where processes are raising interrupts and the OS scheduler is granting processing and access rights to the processes. In the case of the QCOS, the processes are the components from Fig.~\ref{fig:arch}. Overall, the functionality of the QCOS has to be perfectly timed, because the quantum hardware cannot wait, for example, for an optimisation component to be stuck in local minima.

\section{Running the QCOS on a supercomputer}

The aggregated QCOS should be easily executed on a supercomputer. The QCOS problem is difficult, but not extravagantly complex in the sense of computer science theory, and throwing more hardware at it to solve it should be fine~\cite{haigh2018hey}. The advantages of using a supercomputer are that the nodes communicate through very high speed links, message passing is natively supported, and supernodes could be used for each QCOS component.

Supercomputers are designed for high performance and availability. Almost surely these supercomputers would be able to process at the necessary throughput, and for the QCOS, the performance is secondary to the extreme reliability necessary. This reliability can be achieved by a large amount of redundant use of nodes.

Redundant use of nodes becomes necessary when considering the evidence that, in the case of some supercomputers, the reliability consistently degrades down to half by the end of the computation~\cite{rojas2019analyzing}. Most errors in a supercomputer seem to be generated by GPUs~\cite{rojas2019analyzing}, and there is evidence that the mean-time-between-failure (MTBF) of full-system runs might be as low as one day~\cite{ostrouchov2020gpu} -- long quantum computations might fail not because of quantum noise, but because the supercomputer failed. High availability, e.g. 99.98\% (approx. one hour per year downtime), can be achieved by improving the design of the hardware~\cite{ostrouchov2020gpu}, adding automatic network and topology reconfiguration and fault-tolerant routing between the nodes~\cite{zu2024resiliency}.

Even with a fraction of the nodes used for computing, and the other nodes used for increasing the computational reliability to extreme levels, a supercomputer would enable the scalable control of millions of physical qubits, and support the fault-tolerance of the QCOS. Nevertheless, for some tasks, such as the decoding of quantum error-correcting codes (e.g.~\cite{battistel2023real, skoric2023parallel}), the communication latency between a quantum computer and supercomputer might be too large.

The cost of a supercomputer QCOS does not seem exaggerated when considering that the most powerful ones are around \$400 million. It is reasonable to assume that the million-qubit quantum computer will cost more than the supercomputer needed to control it.

\section{Conclusion}

Extremely reliable QCOS will be a necessity. The architecture of a reliable QCOS should start from lessons learned in aerospace engineering and use microkernels, message passing and supercomputers for execution. Nobody has built a simple aggregated QCOS starting from all the freely available software for compiling, optimising and error-correcting quantum circuits. The impossibility of controlling a large scale quantum computer could be rebutted by a supercomputer aggregated QCOS capable of handling the most horrific worst-case scenarios of error-correction.

\section*{Acknowledgements}

This research was developed in part with funding from the Defense Advanced Research Projects Agency [under the Quantum Benchmarking (QB) program under award no. HR00112230006 and HR001121S0026 contracts]. The views, opinions and/or findings expressed are those of the author(s) and should not be interpreted as representing the official views or policies of the Department of Defense or the U.S. Government.

\bibliographystyle{ACM-Reference-Format}
\bibliography{__main}

\end{document}